# Imaging domains in a zero-moment half-metal

K. E. Siewierska[1], N. Teichert[1] and R. Schäfer[2], and J.M.D. Coey[1]

[1]School of Physics & CRANN, Trinity College Dublin, Dublin 2, Ireland
[2]Leibniz Institute for Solid State and Materials Research, Dresden, Germany

We have a choice of methods for examining domains at the surface of a ferromagnet that depend on probing the stray field distribution, but these methods do not work in antiferromagnets or compensated ferrimagnets, which produce no stray field. The discovery of compensated ferrimagnetic half-metals allows for the local magnetization state to be observed directly with polarized light. The example considered here, $Mn_2Ru_xGa$, has two inequivalent but oppositely-aligned Mn sublattices with equal and opposite moments, but only one of them contributes spin polarized conduction electrons at the Fermi energy. The material looks like an antiferromagnet from the outside, but from the point of view of electronic structure it resembles a spin-polarized ferromagnetic metal. The anisotropy axis is perpendicular to the film plane, which allows domains to be imaged directly by polar magneto-optic Kerr effect. The domain structure in a film with a composition of $Mn_2Ru_{0.4}Ga$ has been imaged in a Kerr microscope and hysteresis loops traced. Domains have dimensions of order 20 μm with meandering domain walls and a fractal dimension $D_f = 1.85$. Our results open new direct imaging possibilities for magnetically-ordered materials with no net moment.

*Index Terms*—Kerr microscopy, compensated ferrimagnets, half-metal, $Mn_2Ru_xGa$, antiferromagnetic domain walls

## I. INTRODUCTION

Magnetic domains in antiferromagnets and compensated ferrimagnets are difficult to observe because the materials create no stray field. Methods available to image domains or domain walls in a ferromagnet, at least at the surface of a specimen, include sensing the stray field distribution by atomic force or scanning Hall microscopy, or colloid methods, imaging the magnetic induction $B$ in very thin films by Lorentz transmission electron microscopy, and recording the polarization of light transmitted or reflected from a specimen using the magneto-optic Faraday or Kerr effect. However, none of these methods work in an antiferromagnet. Imaging magnetic domains in materials with no stray field is possible via techniques such as polarized neutron beam tomography [1, 2] or X-ray linear magnetic dichroism at synchrotron light sources [3, 4], or by using auxiliary Fe monolayers and spin polarized scanning tunneling microscopy [5]. Methods depending on magnetostriction [6], optical second-harmonic generation [7, 8] or magnetoelectric effects [9] have also been proposed.

Here we show that it is possible to use the magneto-optic Kerr microscope to image domains directly in a new group of materials that produce no stray field. A class of magnetic material was envisaged by van Leuken and de Groot, which they called a 'half-metallic antiferromagnet' [10]. Like a half metallic ferromagnet [11], it was supposed to be fully spin polarized with a spin gap so that electrons of only one spin are present at the Fermi level, but at the same time it would behave magnetically like an antiferromagnet with compensated magnetisation in the ground state. Such materials are insensitive to external fields and demagnetizing effects, yet electronically they resemble a perfect spin-polarized ferromagnetic metal. However, the material is not a true antiferromagnet because the two sublattices are crystallographically inequivalent; It is a ferrimagnet and the sublattices generally have different temperature dependences [12]. Compensation of the ferrimagnetic zero-moment half

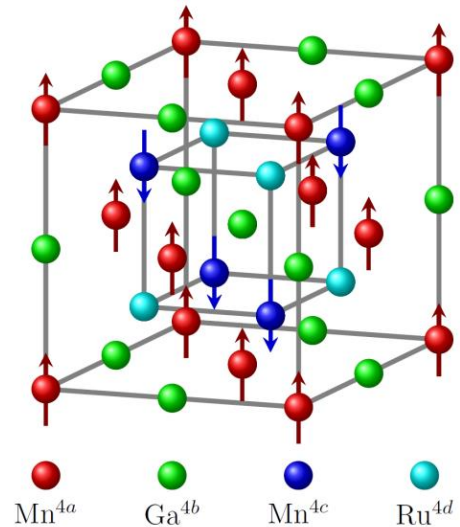

Fig. 1. Model of the inverted L21 crystal structure of $Mn_2Ru_xGa$. The 4d site is only partially occupied by Ru.

metal (ZMHM) will only occur at a specific temperature $T_{comp}$.

The first experimental example of a ZMHM was discovered much later, by Kurt et al. [13], based on electron count in Heusler alloy structures [14]. The compound $Mn_2Ru_xGa$ (MRG) has two equal and oppositely directed but crystallographically-inequivalent, manganese sublattices. The crystal has an inverted $L2_1$ structure as shown in Fig. 1. The Mn atoms at the $4a$ and $4c$ sites form the two sublattices that are responsible for the magnetism of the material. The moments of each sublattice have different temperature dependences, with the $Mn^{4a}$ moment remaining roughly constant up to room temperature and the $Mn^{4c}$ moment decreasing with increasing temperature [15]. This leads to compensation of the net moment at $T_{comp}$ where the moments are equal in magnitude, but opposite in direction. The temperature can be tuned by changing the concentration of Ru, where lower Ru concentration causes a lower $T_{comp}$, but also by substrate induced strain or the presence of Ga antisites [16].

Films of MRG grown on MgO (100) substrates exhibit perpendicular magnetic anisotropy (PMA) due to a small 1% tetragonal elongation induced by biaxial substrate-induced strain. The electronic band structure exhibits a spin gap and it is the $4c$ sublattice that predominantly contributes spin polarized conduction electrons at the Fermi energy. Hence the possibility of imaging the domains of MRG directly by polar magneto-optic Kerr effect (MOKE), thereby opening up new imaging possibilities for domains in magnetically ordered materials with no net moment. In this work we use MOKE to measure the magnetic hysteresis and image the magnetization processes of low-moment MRG films close to compensation point. This serves as a proof-of-concept for imaging magnetic domains in spin polarized, fully compensated ferrimagnets.

## II. EXPERIMENTAL METHODS

Epitaxial thin film of MRG was grown on 10 x 10 mm² MgO (001) substrate by DC magnetron sputtering, using our Shamrock sputtering system. The film was co-sputtered in argon from two 75 mm targets of Mn₂Ga and Ru onto the MgO substrate maintained at 380 °C. The film was capped *in-situ* with a 3 nm layer of AlO$_x$ deposited at room temperature from an Al₂O₃ target in order to prevent oxidation and post annealed *ex-situ* at 350 °C in vacuum for 30 minutes.

The film was deposited with a nominal composition of Mn₂Ru₀.₄Ga and a magnetization < 30 kAm⁻¹ corresponding to a moment < 0.16 μ$_B$ per formula unit. Field and temperature dependent magnetometry data of similar samples are published in Ref. [17]. However, due some Mn evaporation in the post-annealing process, the actual composition is slightly different. The composition Mn₂Ru₀.₄Ga was chosen so that the film has a low enough coercivity to be saturated at room temperature in the 300 mT magnet in the polar Kerr microscope.

A Bruker D8 X-ray diffractometer with a copper tube emitting Kα₁ X-rays with wavelength 154.06 pm and a double bounce Ge [220] monochromator was used to determine the diffraction patterns of the thin films. Low angle X-ray reflectivity (XRR) was measured using a Panalytical X'Pert

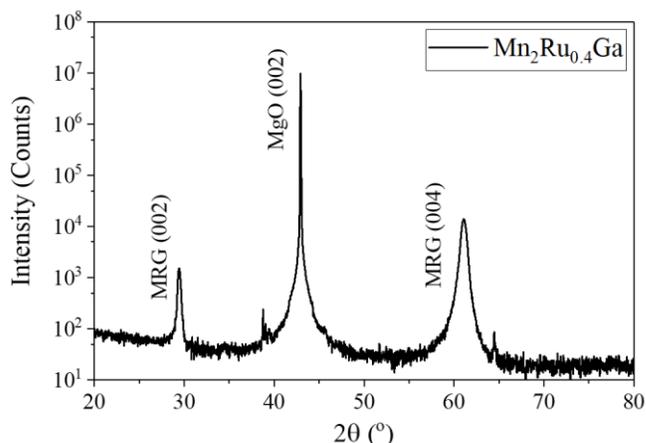

Fig. 2. X-ray diffraction pattern of a film of Mn₂Ru₀.₄Ga on MgO.

Pro diffractometer, and thickness was found by fitting the interference pattern using X'Pert Reflectivity software with a least square fit.

Anomalous Hall effect was measured using the four-point Van-der-Pauw-method with indium contacts and an applied current of 5 mA.

Atomic force microscopy was performed using DI Multi-Mode Atomic Force Microscope in contact mode.

For Kerr imaging, a wide-field Kerr microscope with a 10×/0.25 objective lens was used. All loops have been measured with polar sensitivity and a control feature which readjusts the analyzer position with the help of a reference mirror in order to compensate the Faraday rotation in the objective lenses caused by the magnetic field applied perpendicular to the film [18]. The domain contrast was optimized by the means of an analyzer and rotatable compensator (quarter wave plate) [19]. Creeping behavior was observed on changing the magnetic field, so the images were recorded after the domain structure had stabilized.

The greyscale domain images were converted to binary images using ImageJ software and the fractal dimension of the domains were calculated by the box-counting method using Fractalyse [20] software.

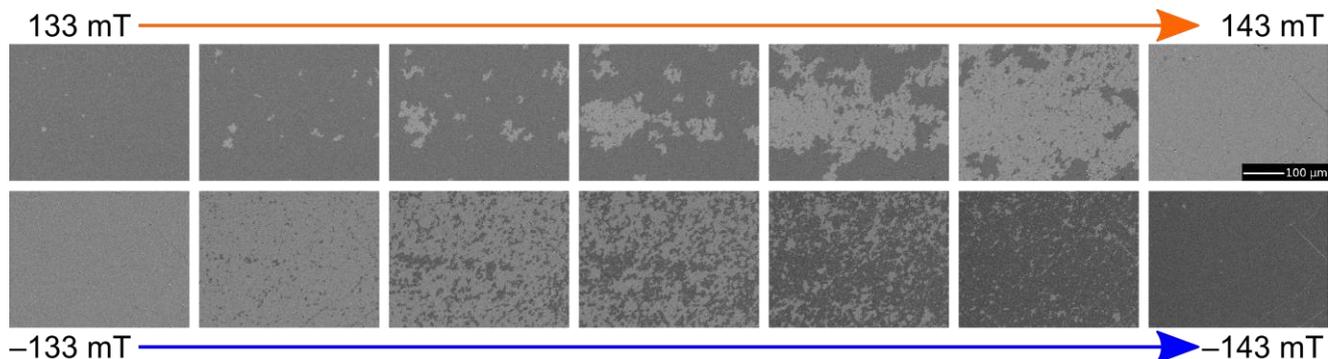

Fig. 3. Kerr imaging of the magnetization reversal process of nearly compensated Mn₂Ru₀.₄Ga with out-of-plane easy axis. The contrast changes from dark (light) to light (dark) with increasing positive (negative) field caused by nucleation and growth of magnetic domains during the switching process.

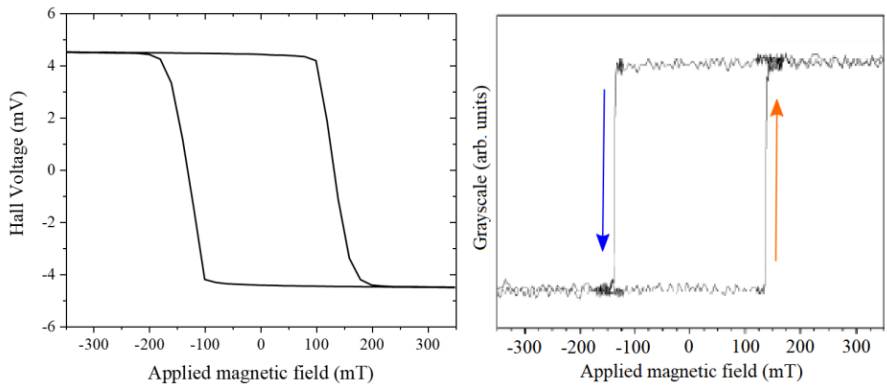

Fig. 4. Hysteresis loops of $Mn_2Ru_{0.4}Ga$ obtained via (a) anomalous Hall effect and (b) Kerr microscopy. The sharp switching observed by MOKE on a submillimeter sized image, contrasts with the broader switch measured by Hall effect on a blanket $10 \times 10$ mm$^2$ film in the van-der-Pauw configuration.

## III. RESULTS AND DISCUSSION

The x-ray diffraction pattern of the $Mn_2Ru_{0.4}Ga$ thin film shows (002) and (004) reflections, together with peaks from the MgO substrate. The ratio of (002) and (004) peak intensities (I(002)/I(004)) was 0.06 and this small value is indicative of a high degree of atomic order in the inverted L2$_1$ Heusler structure [21]. The c lattice parameter was calculated to be 606.4 pm hence (c - a)/a is approximately 1% indicating the percentage of tetragonal distortion of the cubic unit cell [13]. A least-squares fit to the XRR pattern yielded a thickness $t$ for the MRG thin film of 40.2 nm. Atomic force microscopy of the film shows an rms surface roughness of 1.3 nm.

Polar Kerr microscopy was used to follow the magnetic switching process of the perpendicular $Mn_2Ru_{0.4}Ga$ film. The domain images in a 400 x 300 μm$^2$ window are presented in Fig. 3 as a function of applied field around the negative and positive coercive fields. The magnetic hysteresis loops obtained by measurement of the anomalous Hall effect and analysis of the Kerr microscopy images are compared in Fig. 4(a) and 4(b). Switching fields of both measurements coincide, but the switching process observed by MOKE is sharper. We attribute this to slight inhomogeneities in the film that are less visible in MOKE due to the smaller probing area of the measurement ($10 \times 10$ mm$^2$ blanket film for the Hall measurement vs a submillimeter sized image for MOKE). According to Fig. 4 (b), the coercive field $\mu_0 H_C$ = 137 mT and the onset of magnetization switching is at $\mu_0 H_{SW}$ = 133 mT. Domain creep is observed in the vicinity of the coercive field. The arrows indicate the direction of magnetization change. The emerging domains (bright contrast) are seen to have irregular outlines, with dimensions of order 20 μm. They appear to nucleate randomly. With increasing negative field, the reverse domains grow, and more new ones appear until the entire film has switched. We anticipate that defects or surface roughness of the film may play a part in pinning domain walls. The evolution of the domain pattern around negative $H_C$ (lower panel in Fig. 3, left to right) is similar and also involves nucleation and growth of irregularly shaped domains (dark contrast) with a fractal structure. The same domain patterns are not reproduced on field cycling to saturation. They do not nucleate in the same places, and nucleation does not appear to be primarily controlled by a set of nonmagnetic defects.

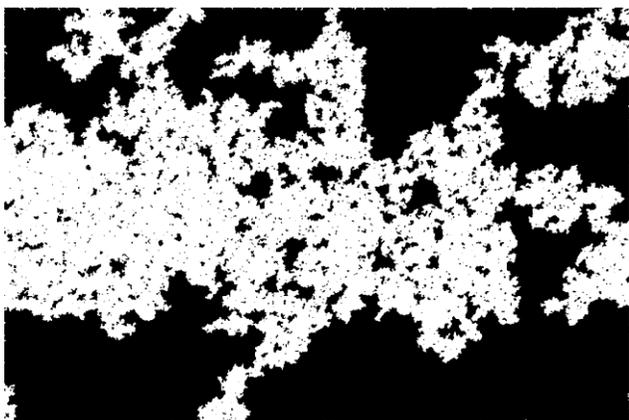

Fig. 5. Fractal analysis. Fifth image (from left) of the top panel of Fig. 3, converted from greyscale to binary. The image has a fractal dimension D$_f$ = 1.85.

Fractal-like magnetic domains have previously been observed in thin films [22, 23] and at surfaces [24] of strong uniaxial ferromagnets. Fractal antiferromagnetic domains have been reported in NiO using spectroscopic photoemission and low energy electron microscopy [25]. Fig. 5 reproduces the fifth image of the top panel of Fig. 3 after binary conversion using ImageJ. The fractal dimension of the domains determined by the box-counting method is $D_f$ = 1.85. The dimension extracted theoretically for two dimensional percolating clusters is 91/48 = 1.896 [26].

The irregular, fractal-like outlines of the domains suggest that the domain wall energy (per unit area) may be smaller than in typical ferromagnets, or else the antiferromagnetic domain walls may be topologically protected.

## IV. CONCLUSIONS

Imaging of magnetic domains in low moment ferrimagnetic thin films with high spin polarization via magneto-optic Kerr effect has been demonstrated. Our approach offers an opportunity to investigate the domain structure of ZMHM materials at $T_{comp}$ by Kerr microscopy. The method can be extended to provide insight into domain dynamics in the absence of stray field. Due to the uniaxial anisotropy in MRG thin films this material shows a similar magnetic structure to a uniaxial antiferromagnet and therefore, it is a candidate for investigating nucleation and dynamics of the elusive 180° antiferromagnetic domain walls, and also to study spin-flop transitions on a microscopic scale. Domains can also be observed at compensation by varying the sample temperature.

## ACKNOWLEDGMENT

We are grateful to Siddhartha Sen for helpful discussions. The work of K. E. Siewierska was supported by the Irish Research Council under Grant GOIPG/2016/308. N. Teichert would like to acknowledge funding from the European Union's Horizon 2020 research and innovation programme under the Marie Skłodowska-Curie EDGE grant agreement No 713567. The work form part of the SFI-supported ZEMS project, grant number 16/IA/4534

## V. REFERENCES

[1] M. Schlenker and J. Baruchel, "Neutron techniques for the observation of ferro- and antiferromagnetic domains," *Journal of Applied Physics*, vol. 49, no. 3, pp. 1996–2001, 1978.

[2] M. Schlenker, J. Baruchel, J. F. Pétroff, and W. B. Yelon, "Observation of subgrain boundaries and dislocations by neutron diffraction topography," *Appl. Phys. Lett.*, vol. 25, no. 7, pp. 382–384, 1974.

[3] M. J. Grzybowski et al., "Imaging Current-Induced Switching of Antiferromagnetic Domains in CuMnAs," (eng), *Phys. Rev. Lett.*, vol. 118, no. 5, p. 57701, 2017.

[4] Nolting et al., "Direct observation of the alignment of ferromagnetic spins by antiferromagnetic spins," *Nature*, vol. 405, no. 6788, pp. 767–769, 2000

[5] M. Bode et al., "Atomic spin structure of antiferromagnetic domain walls," *Nat. Mater.* vol. 5, no. 6, pp. 477–481, 2006.

[6] B. K. Tanner, "Antiferromagnetic domains," *Contemporary Physics*, vol. 20, no. 2, pp. 187–210, 1979.

[7] Fiebig, Fröhlich, Krichevtsov, and Pisarev, "Second harmonic generation and magnetic-dipole-electric-dipole interference in antiferromagnetic $Cr_2O_3$,", *Phys. Rev. Lett.*, vol. 73, no. 15, pp. 2127–2130, 1994.

[8] M. Fiebig, D. Fröhlich, G. Sluyterman v. L., and R. V. Pisarev, "Domain topography of antiferromagnetic $Cr_2O_3$ by second-harmonic generation," *Appl. Phys. Lett.* vol. 66, no. 21, pp. 2906–2908, 1995.

[9] P. Schoenherr et al., "Magnetoelectric Force Microscopy on Antiferromagnetic 180° Domains in $Cr_2O_3$," *Materials (Basel)*, vol. 10, no. 9, 2017.

[10] H. van Leuken and R. A. de Groot, "Half-metallic antiferromagnets," *Phys. Rev. Lett.*, vol. 74, no. 7, pp. 1171–1173, 1995.

[11] R. A. de Groot, F. M. Mueller, P. G. van Engen, and K. H. J. Buschow, "New Class of Materials: Half-Metallic Ferromagnets," *Phys. Rev. Lett.* vol. 50, no. 25, pp. 2024–2027, 1983.

[12] E. Şaşıoğlu, "Nonzero macroscopic magnetization in half-metallic antiferromagnets at finite temperatures," *Phys. Rev. B*, vol. 79, no. 10, p. 100406, 2009.

[13] H. Kurt et al., "Cubic $Mn_2Ga$ thin films: Crossing the spin gap with ruthenium," *Phys. Rev. Lett.* vol. 112, no. 2, p. 27201, 2014.

[14] T. Graf, C. Felser, and S. S.P. Parkin, "Simple rules for the understanding of Heusler compounds," *Progress in Solid State Chemistry*, vol. 39, no. 1, pp. 1–50, 2011.

[15] D. Betto et al., "Site-specific magnetism of half-metallic $Mn_2Ru_xGa$ thin films determined by x-ray absorption spectroscopy," *Phys. Rev. B*, vol. 91, no. 9, p. 094112, 2015.

[16] M. Žic et al., "Designing a fully compensated half-metallic ferrimagnet," *Phys. Rev. B*, vol. 93, no. 14, p.140402, 2016.

[17] K. E. Siewierska et al., "Study of the Effect of Annealing on the Properties of $Mn_2Ru_xGa$ Thin Films," *IEEE Trans. Magn.*, vol. 53, no. 11, p. 7929426, 2017.

[18] I. V. Soldatov and R. Schäfer, "Advanced MOKE magnetometry in wide-field Kerr-microscopy," *Journal of Applied Physics*, vol. 122, no. 15, p. 153906, 2017.

[19] R. Schäfer, "Investigation of Domains and Dynamics of Domain Walls by the Magneto-optical Kerr-effect," in *Handbook of Magnetism and Advanced Magnetic Materials*, H. Kronmüller and S. Parkin, Eds., Chichester, UK: John Wiley & Sons, Ltd, 2007.

[20] http://www.fractalyse.org/.

[21] D. Betto et al., "The zero-moment half metal: How could it change spin electronics?" *AIP Advances*, vol. 6, no. 5, p. 55601, 2016.

[22] D.-H. Kim, Y.-C. Cho, S.-B. Choe, and S.-C. Shin, "Correlation between fractal dimension and reversal behavior of magnetic domain in Co/Pd nanomultilayers," *Appl. Phys. Lett.*, vol. 82, no. 21, pp. 3698–3700, 2003.

[23] D. Navas et al., "Microscopic reversal magnetization mechanisms in CoCrPt thin films with perpendicular magnetic anisotropy: Fractal structure versus labyrinth stripe domains," *Phys. Rev. B*, vol. 96, no. 18, p.189403, 2017.

[24] A. Kreyssig et al., "Probing fractal magnetic domains on multiple length scales in $Nd_2Fe_{14}B$," *Phys. Rev. Lett.*, vol. 102, no. 4, p. 47204, 2009.

[25] J. Das and K. S.R. Menon, "On the evolution of antiferromagnetic nanodomains in NiO thin films: A LEEM study," *Journal of Magnetism and Magnetic Materials*, vol. 449, pp. 415–422, 2018.

[26] D. He et al., "Two-dimensional percolation and cluster structure of the random packing of binary disks," *Phys. Rev. E*, vol 65, p. 061304, 2002.